\documentclass[ article , amsmath ,amssymb , aps , onecolumn ]{revtex4-2}

\usepackage{amssymb}
\usepackage[caption=false]{subfig}
\usepackage{amssymb}
\usepackage{amsmath}
\usepackage{float}
\usepackage[caption = false]{subfig}
\usepackage[final]{graphicx}
\usepackage{verbatim}
\usepackage[normalem]{ulem}
\usepackage[utf8]{inputenc}
\usepackage{caption}

\begin{document}

\title{Constraining scalarization in scalar-Gauss-Bonnet gravity through binary pulsars}

\author{Victor~I.~Danchev$^1$}
\homepage{Electronic address: \href{vidanchev@uni-sofia.bg}{vidanchev@uni-sofia.bg}}
\author{Daniela~D.~Doneva$^{2,3}$}
\homepage{Electronic address: \href{daniela.doneva@uni-tuebingen.de}{daniela.doneva@uni-tuebingen.de}}
\author{Stoytcho~S.~Yazadjiev$^{1,2,4}$}
\homepage{Electronic address: \href{yazad@phys.uni-sofia.bg}{yazad@phys.uni-sofia.bg}}
\affiliation{$^1$Department of Theoretical Physics, Faculty of Physics, Sofia University, Sofia 1164, Bulgaria}
\affiliation{$^2$Theoretical Astrophysics, Eberhard Karls University of Tübingen, Tübingen 72076, Germany}
\affiliation{$^3$INRNE -- Bulgarian Academy of Sciences, 1784 Sofia, Bulgaria}
\affiliation{$^4$Institute of Mathematics and Informatics, Bulgarian Academy of Sciences, Acad. G. Bonchev Street 8, Sofia 1113, Bulgaria}

\begin{abstract}
In the present paper we derive strong constrains on scalarization in scalar-Gauss-Bonnet (sGB) gravity using observations of pulsars in close binary systems. 
Since scalarized neutron stars carry a nonzero scalar change, they emit scalar dipole radiation while inspiraling which speeds up the orbital decay. 
The observations support the conjecture that such radiation is either absent or very small for the observed binary pulsars. 
Using this, we determine the allowed range of parameters for sGB gravity. We also transfer the derived constraints to black holes in sGB gravity. It turns out that the maximum mass of a scalarized black hole can not exceed roughly ten to twenty solar masses, depending on the initial assumptions we make for the nuclear matter equations of state. The black hole scalar charge on the other hand can reach relatively large values that are potentially observable.      
\end{abstract}

\maketitle

\section{Introduction}\label{introduction}
The strong field regime of gravity is still a not well-explored area that leaves a lot of freedom for modifications beyond general relativity (GR) \cite{Berti:2015itd}. 
Historically, one of the first observations that could constrain this regime were the pulsars in close binary systems \cite{Damour:1996ke,Yunes:2013dva,Seymour:2018bce}. 
More specifically, one can observe the shrinking of the orbit due to the emission of gravitational radiation. 
The observations show that this orbital decay fits very well the predictions of GR that constrained severely some theories of gravity which predict an additional channel of energy loss \cite{Freire:2012mg,Antoniadis:2013pzd,Shao:2017gwu,Seymour:2020yle,Shao:2020fka}. 
A prominent example in this context are the theories admitting scalarization and the scalar-tensor Damour-Esposito-Farese (DEF) model in particular. 
Such theories are characterized by field equations that are perturbatively equivalent to GR in the weak field regime while allowing for strong deviations from Einstein's theory of gravity in the realm of large spacetime curvature. 
The presence of nonzero scalar charge of the neutron stars in the DEF model leads to the emission of scalar gravitational radiation that is an additional channel of energy loss and causes the binary system to inspiral faster compared to GR. 
Since this is not actually observed for pulsars in close binary systems, strong constraints were imposed on the DEF model \cite{Damour:1996ke,Damour:1998jk,Freire:2012mg,Antoniadis:2013pzd,Shao:2017gwu,Voisin:2020lqi,Kramer:2021jcw}. 
Due to the extremely high accuracy of the pulsar timing observations, it turns out that these constraints can not be easily improved by the observations of other gravitational wave sources such as binary neutron star mergers where indeed the gravitational wave signal from the inspiral is much stronger but the accuracy is far inferior \cite{Damour:1998jk,Sampson:2014qqa}. 
Thus the observations of pulsars in close binary systems still remain among the ultimate tools to test the strong-field regime of gravity.

A class of alternative theories of gravity where scalarization was discovered a few years ago are the Gauss-Bonnet (GB) theories \cite{Kanti:1995vq}.
They are perhaps the most prominent subclass of quadratic theories of gravity defined by the addition of all possible geometrical invariants of second-order to the theory action in addition to the classical Einstein-Hilbert term \cite{Stelle:1976gc}. 
A strong motivation behind these types of modifications of Einstein's theory comes from the attempts to quantize gravity. 
What makes Gauss-Bonnet theories special is the fact that the field equations are of second-order lacking ghosts and they can be viewed also as a subclass of the Horndeski theories. 

Compact objects in Gauss-Bonnet theories were first considered in the context of the so-called Einstein dilaton Gauss-Bonnet (EdGB) gravity where the coupling between the scalar field and the Gauss-Bonnet invariant is a linear function of the scalar field or an exponent of a linear function \cite{Mignemi:1992nt,Yunes:2011we,Pani:2011xm,Pani:2011gy,Sotiriou:2013qea,Sotiriou:2014pfa,Ayzenberg:2014aka,Kleihaus:2011tg,Kleihaus:2014lba,Kleihaus:2016dui}. 
The studies showed that in this case the scalar charge and thus the scalar dipole radiation are either zero or negligible and hence the binary pulsar observations can not impose strong constraints on the theory \cite{Yagi:2015oca}. 
This is not true, though, for classes of Gauss-Bonnet theories allowing for scalarization, that we will call scalar-Gauss-Bonnet (sGB) theories, for which the scalar charge can be large \cite{Doneva:2017bvd,Silva:2017uqg,Antoniou:2017acq}. 

Even though scalarization in sGB gravity originally attracted a lot of attention in the context of black holes, it is possible to scalarize also neutron stars \cite{Silva:2017uqg,Doneva:2017duq,Xu:2021kfh}.
Thus one can impose constraints on the theory through binary pulsar observations which is the goal of the present paper. 
This is important because even though scalarization in sGB gravity brought a lot of excitement, little is known about the astrophysical implications. 
In particular, we constrain the possible range of values for the sGB gravity parameters through Markov Chain Monte-Carlo (MCMC) methods applied to the observation of the most suitable pulsar-white dwarf systems that are used also to constrain scalarization in the DEF model \cite{Shao:2017gwu}. 
This has been performed by comparing the predicted effects due to scalar dipole radiation on the binary's orbital period evolution (which is lacking for the pure GR case) with the observed period evolution, considering the experimental measurement errors.

The paper is structured as follows. In Sec. II we briefly review the theory behind sGB gravity and building scalarized neutron star models in this theory. 
In Sec. III the methodology for constraining the theory through observations of pulsars in binary systems is described. 
The code setup and the results are discussed in Sec. IV. 
Sec. V is devoted to transferring these constraints to the black hole scalarization. 
The paper ends with Conclusions.

\section{Gravitational theory formulation}\label{theory}
We are working within the framework of sGB theory having the following action
\begin{eqnarray}\label{action}
    S = \frac{ 1 }{ 16\pi G_{*} } \int d^4x \sqrt{ - g } \left[ R - 2 \nabla_{ \mu } \varphi \nabla^{ \mu } \varphi + \lambda^2 f( \varphi ) \mathcal{ R }_{GB}^2 \right] + S_{ \mathrm{ matter } }( g_{ \mu \nu } , \chi ),
\end{eqnarray}
where $R$ is the Ricci scalar and $\nabla_{\mu}$ is the covariant derivative, both with respect to the spacetime metric $g_{\mu \nu}$.
The theory choice is made by setting the coupling function for the scalar field $f(\varphi)$ and the GB coupling constant $\lambda$ which has the dimension of length. 
Lastly, the Gauss-Bonnet invariant is defined as $ \mathcal{ R }^2_{ GB } = R^2 - 4 R_{ \mu \nu } R^{ \mu \nu } + R_{ \mu \nu \alpha \beta } R^{ \mu \nu \alpha \beta }$ with respect to the spacetime Ricci and Riemann tensors.
The field equations corresponding to that action are
\begin{eqnarray}\label{field_eq_metric}
    R_{ \mu \nu } - \frac{ 1 }{ 2 } R g_{ \mu \nu } + \Gamma_{ \mu \nu } &=& 2 \nabla_{ \mu } \varphi \nabla_{ \nu } \varphi - g_{ \mu \nu } \nabla_{ \alpha } \varphi \nabla^{ \alpha } \varphi + 8 \pi G_{*} T_{ \mu \nu }^{ \mathrm{ matter } } \\ \label{field_eq_phi}
    \nabla_{ \alpha } \nabla^{ \alpha } \varphi &=& - \frac{ \lambda^2 }{ 4 } \frac{ df( \varphi ) }{ d \varphi } \mathcal{ R }^2_{ GB },
\end{eqnarray}
where $T_{\mu \nu}^{\mathrm{matter}}$ is the matter energy-momentum tensor while $ \Gamma_{ \mu \nu } $ is defined as
\begin{eqnarray}\label{gamma_def}
    \Gamma_{ \mu \nu } =& - R( \nabla_{ \mu } \Psi_{ \nu } + \nabla_{ \nu } \Psi_{ \mu } ) - 4 \nabla^{ \alpha } \Psi_{ \alpha } \left( R_{ \mu \nu } - \frac{ 1 }{ 2 } R g_{ \mu \nu } \right) + 4 R_{ \mu \alpha } \nabla^{ \alpha } \Psi_{ \nu } + 4 R_{ \nu \alpha } \nabla^{ \alpha } \Psi_{ \mu } \nonumber \\
    &- 4 g_{ \mu \nu } R^{ \alpha \beta } \nabla_{ \alpha } \Psi_{ \beta } + 4 R^{ \beta }_{ \mu \alpha \nu } \nabla^{ \alpha } \Psi_{ \beta },
\end{eqnarray}
and we have defined
\begin{eqnarray}\label{psi_def}
    \Psi_{ \mu } = \lambda^2 \frac{ df( \varphi ) }{ d \varphi } \nabla_{ \mu } \varphi.
\end{eqnarray}
By applying the field equations to the contracted Bianchi identities, one can show that the matter energy-momentum tensor satisfies the conservation law 
\begin{eqnarray}\label{EM_conservation}
    \nabla^{ \mu }T_{ \mu \nu }^{ \mathrm{ matter } } = 0.
\end{eqnarray}
In the present paper we are studying static and spherically symmetric neutron stars in sGB gravity so the spacetime metric is given by the following ansatz 
\begin{eqnarray}
    dS^2 = - e^{ 2 \Phi( r ) }dt^2 + e^{ 2 \Lambda( r ) }dr^2 + r^2( d\theta^2 + \sin^2{ \theta } d\phi^2 ).
\end{eqnarray}
The matter source is taken as perfect fluid $ T_{ \mu \nu }^{ \mathrm{ matter } } = ( \rho + P ) u_{ \mu } u_{ \nu } + P g_{ \mu \nu } $ where $ \rho $, $ P $ and $ u^{ \mu } $ are respectively the energy density, the pressure and the 4-velocity of the fluid.
Lastly, we require the perfect fluid and the scalar field to respect the staticity and spherical symmetry of the spacetime which leads us to the following dimensionally reduced field equations
\begin{eqnarray}\label{red_field_eq_1}
    \frac{ 2 }{ r }\left[ 1 + \frac{ 2 }{ r }( 1 - 3 e^{ - 2 \Lambda } )\Psi_r \right] \frac{ d\Lambda }{ dr } + \frac{ ( e^{ 2 \Lambda } - 1 ) }{ r^2 } - \frac{ 4 }{ r^2 }( 1 - e^{ - 2 \Lambda } ) \frac{ d \Psi_r }{ dr } - \left( \frac{ d\varphi }{ dr } \right)^2 &=& 8\pi G_{*} \rho e^{ 2 \Lambda }, \\ \label{red_field_eq_2}
    \frac{ 2 }{ r }\left[ 1 + \frac{ 2 }{ r }( 1 - 3 e^{ - 2 \Lambda } )\Psi_r \right] \frac{ d\Phi }{ dr } - \frac{ ( e^{ 2 \Lambda } - 1 ) }{ r^2 } - \left( \frac{ d\varphi }{ dr } \right)^2 &=& 8\pi G_{*} P e^{ 2 \Lambda }, \\ \label{red_field_eq_3}
    \frac{ d^2 \Phi }{ dr^2 } + \left( \frac{ d\Phi }{ dr } + \frac{ 1 }{ r } \right) \left( \frac{ d\Phi }{ dr } - \frac{ d\Lambda }{ dr } \right) + \frac{ 4e^{ - 2\Lambda } }{r}\left[ 3 \frac{ d\Phi }{ dr }\frac{ d\Lambda }{ dr } - \frac{ d^2 \Phi }{ dr^2 } - \left( \frac{ d\Phi }{ dr } \right)^2 \right] \Phi_r \nonumber \\
    - \frac{ 4 e^{ - 2\Lambda } }{ r }\frac{ d\Phi }{ dr }\frac{ d\Psi_r }{ dr } + \left( \frac{ d\varphi }{ dr } \right)^2 &=& 8\pi G_{*} P e^{ 2\Lambda }, \\ \label{red_field_eq_4}
    \frac{ d^2 \varphi }{ dr^2 } + \left( \frac{ d\Phi }{ dr } - \frac{ d\Lambda }{ dr } + \frac{ 2 }{ r } \right)\frac{ d\varphi }{ dr } \nonumber \\
    - \frac{ 2 \lambda^2 }{ r^2 }\frac{ df( \varphi ) }{ d\phi }\left\{ ( 1 - e^{ - 2 \Lambda } ) \left[ \frac{ d^2 \Phi }{ dr^2 } + \frac{ d\Phi }{ dr }\left( \frac{ d\Phi }{ dr } - \frac{ d\Lambda }{ dr } \right) \right] + 2e^{ - 2\Lambda } \frac{ d\Phi }{ dr }\frac{ d\Lambda }{ dr } \right\} &=& 0,
\end{eqnarray}
where from eq. \eqref{psi_def} we have
\begin{eqnarray}
    \Psi_r = \lambda^2 \frac{ df( \varphi ) }{ d\varphi }\frac{ d\varphi }{ dr }.
\end{eqnarray}
Additionally, the hydrostatic equilibrium equation takes the form
\begin{eqnarray}
    \frac{ dP }{ dr } = - ( \rho + P )\frac{ d\Phi }{ dr },
\end{eqnarray}
which follows from \eqref{EM_conservation} applied for a perfect fluid.

The regularity at the center and asymptotic flatness impose the following boundary conditions 
\begin{eqnarray}
\Lambda|_{r\rightarrow 0} \rightarrow 0, \;\; \left.\frac{d\Phi}{dr}\right|_{r\rightarrow 0} \rightarrow 0, \;\; \left.\frac{d\varphi}{dr}\right|_{r\rightarrow 0} \rightarrow 0\;\;.   \label{eq:BC_r0}
\end{eqnarray} 
and
\begin{eqnarray}
\Lambda|_{r\rightarrow\infty} \rightarrow 0,\;\; \Phi|_{r\rightarrow\infty} \rightarrow 0, \;\; \varphi|_{r\rightarrow\infty} \rightarrow 0\;\;.   \label{eq:BC_rinf}
\end{eqnarray} 
The leading order asymptotic of the scalar field  at infinity has the form 
\begin{equation}\label{eq:Scalar_Charge}
\varphi\approx \frac{D}{r} + O(1/r^2).
\end{equation}
where the scalar charge $D$ is a constant directly related to the gravitational wave emission in close binary systems as we will discuss below.

The existence and stability of scalarized neutron star solutions has been studied by two of the present authors in \cite{Doneva:2017duq}. In the present work we are solving equations \eqref{red_field_eq_1}--\eqref{red_field_eq_4} with the following commonly employed coupling function
\begin{eqnarray}\label{eq:coupling}
    f( \varphi ) = \frac{ \epsilon }{ 2\beta }[ 1 - \exp( - \beta \varphi^2 )]
\end{eqnarray}
with $\beta > 0$ and $\epsilon = \pm 1$. 
While for both $\epsilon = \pm 1$ scalarized neutron stars solutions exist, only for positive $\epsilon$ static black holes can scalarize \cite{Doneva:2017bvd}. 
The minus sign of $\epsilon$ is associated with the so-called spin-induced scalarization \cite{Dima:2020yac,Hod:2020jjy,Doneva:2020nbb,Berti:2020kgk,Herdeiro:2020wei,Doneva:2020kfv} where the development of scalar field is observed only for sufficiently rapidly rotating neutron stars. 
In order to obtain neutron star solutions for different $\beta$ and $\lambda$ values corresponding to one of the partners in a Neutron Star--White Dwarf (NS--WD) binary system we follow closely the methodology described in \cite{Doneva:2017duq}. 
The rationale and constraints that can be imposed from observations of such systems are explored in the following section.
As already shown in \cite{Doneva:2017duq}, the considered theory approaches standard GR as $\beta$ tends to infinity. Therefore, to avoid a growing probability function for larger values in the parameters space, we have reparametrized \eqref{eq:coupling} by the inverse parameter $\kappa \equiv \beta^{-1}$ for all numerical computations in the remainder of the paper.

The coupling function choice \eqref{eq:coupling} is by no means unique. Other couplings have also been used in the literature with the most prominent example being $f( \varphi )=\varphi^2$ \cite{Silva:2017uqg} that matches \eqref{eq:coupling} for very small scalar fields. 
Even though such quadratic coupling is the simplest one admitting scalarization, it leads to unstable scalarized black holes \cite{Blazquez-Salcedo:2018jnn} and the behavior of the scalarized neutron star branches look also somewhat peculiar compared to the exponential coupling \cite{Xu:2021kfh}. 
This can be cured (at least for black holes) by adding higher order $\varphi$ terms in the coupling function \cite{Minamitsuji:2018xde,Silva:2018qhn} or a self-interacting scalar field potential \cite{Macedo:2019sem}. 
From a numerical point of view, though, the exponential coupling is much more preferred when heavier simulations such as binary merger \cite{Ripley:2020vpk} or stellar core-collapse \cite{Kuan:2021lol} are performed, as well as for rotating black hole solutions \cite{Cunha:2019dwb,Herdeiro:2020wei}. 
Since our goal is not only to derive the constraints from binary pulsar experiments but also to apply them to constraint the properties of astrophysically relevant scalarized black holes, it is of great importance to work with a coupling function that leads to well behaved branches of solutions that is secured by the choice \eqref{eq:coupling}. 
Moreover, as the experience from other studies shows \cite{Doneva:2018rou,East:2021bqk}, if we consider a modified coupling that still leads  to well-behaved branches of solutions, the final results will remain qualitatively the same.

\section{Constraints from binary pulsars observation}\label{binaries_theory}
As discussed in the introduction, the pulsars in close binary systems are a very useful tool to probe the strong field regime of gravity due to the possibility for a precise measurement of the orbital decay due to gravitational wave emission. 
We are focusing our investigation on a state-of-the-art observations of a set of three neutron star-white dwarf (NS-WD) binaries listed in Table \ref{table_of_parameters}, following the MCMC strategy as formulated by \cite{Shao:2017gwu}, and derive constraints in the $\kappa$ -- $\lambda$ space. 
These pairs are J0348+0432, J1012+5307 and J2222--0137, and the numerical values of the parameters for them are outlined in the table. 
We should note that the considered binary pulsars are rotating with a spin frequency within the range 30-200 Hz. 
This is still well within the slow rotation regime where the static spherically symmetric solutions are a very good approximation. This is especially true for the stellar mass and scalar charge where the rotational corrections are already of second order with respect to the spin. For this reason we will consider static spherically symmetric scalarized neutron stars as an good approximation to the observed pulsars, similar to \cite{Shao:2017gwu}.

\begin{table}[h!]
\begin{center}
\begin{tabular}{ |c|c|c|c| } 
 \hline
 Quantity & J0348+0432 values & J1012+5307 values & J2222--0137 values \\
 \hline
 Orbital Period ($P_b$) in days & 0.102424062722 ± $7 \times 10^{-12}$ & 0.60467271355 ± $3 \times 10^{-11} $ & 2.445759995471 ± $6 \times 10^{-12}$ \\ 
 Eccentricity ($e$) & $2.6 \times 10^{-6} \pm 9 \times 10^{-7}$ & $ 1.2 \times 10^{-6} \pm 3 \times 10^{-7} $ & $ 3.8092 \times 10^{-4} \pm 4 \times 10^{-8} $ \\ 
 Intrinsic $\dot{P}^{\mathrm{int}}_b$ in ($\mathrm{fs.s^{-1}}$) & -274 ± 45 & -2.1 ± 8.6 & -10 ± 8 \\ 
 NS to WD mass ratio $q \equiv m_p/m_c $ & 11.70 ± 0.13 & 10.44 ± 0.11 & n/a \\
 Pulsar mass $m_p^{\mathrm{obs}}$ in $M_{\odot}$ & *2.0065 $^{+0.0755}_{-0.0570}$ & *1.72 $^{+0.18}_{-0.17}$ & 1.81 ± 0.03 \\
 Observed WD mass $m_c^{\mathrm{obs}}$ in $M_{\odot}$ & 0.1715 $^{+0.0045}_{-0.0030}$ & 0.165 ± 0.015 & 1.312 ± 0.009 \\
 \hline
\end{tabular}
\caption{Physical parameters for the NS--WD pairs used for the inference of constraints. Values are obtained from \cite{Antoniadis:2013pzd, Lazaridis:2009kq, Guo:2021, MataSanchez:2020pys, Ding:2020sig} where the observed time derivatives are corrected using the Galactic potential of \cite{McMillan2017}. Values marked with * are not directly observed but derived from the remaining measurements.}
\label{table_of_parameters}
\end{center}
\end{table}

As a matter of fact initially we had considered more NS--WD binaries but based on the specifics of the problem, that has some important qualitative differences compared to the DEF model \cite{Shao:2017gwu}, we have kept only these three systems that are actually the three most massive pulsars used in \cite{Shao:2017gwu}. 
The higher masses make them well fitted for constraining the Gauss-Bonnet theory with J0348+0432 being the most suitable for the following reason. 
As seen in \cite{Doneva:2017duq}, for greater $\lambda$ the bifurcation point from the GR branch will appear at lower masses.
Therefore, less massive neutron stars will generally provide constraints on $\lambda$ starting from larger $\lambda$ values and will not contribute to a distribution where constraints from more massive stars are considered. 
This will be seen in the numerical results for these three NS--WD pairs in Sec. \ref{results_1}.
Since up to now J0348+0432 is the most massive pulsar with a well measured orbital decay, it provides the best constraints.
The same methodology can be readily applied to future NS--WD pairs with higher pulsar mass but as discussed below, if we assume that the maximum neutron star mass is close to two solar masses, it is expected that other pulsars can improve the constraints only slightly.

A posterior distribution of the parameters is obtained by integration over a set of distributions known as the prior, likelihood and model evidence based on Bayes' theorem. 
Given a hypothesis $\mathcal{H}$ (which is the sGB gravity for certain $\kappa$ and $\lambda$ values in our case) and a data collection $\mathcal{D}$, the marginalized posterior distribution of the parameters is given by 
\begin{eqnarray}\label{Bayes_Eq}
    P( \kappa , \lambda | \mathcal{D}, \mathcal{H}, \mathcal{I} ) = \int \frac{ P( \mathcal{D}|\kappa , \lambda , \mathbf{ \Theta } , \mathcal{H} , \mathcal{I} ) P( \kappa , \lambda , \mathbf{ \Theta } | \mathcal{H} , \mathcal{I} ) }{ P( \mathcal{D} | \mathcal{H}, \mathcal{I} ) } d\mathbf{ \Theta },
\end{eqnarray}
where $\mathcal{I}$ is a collective notation for all prior background knowledge and $\mathbf{ \Theta }$ corresponds to all other unknown parameters except for ($\kappa ,\lambda$).
These are integrated over to obtain the marginalized posterior distribution in terms of just ($\kappa,\lambda$).
Furthermore, in \eqref{Bayes_Eq}, $P( \kappa , \lambda , \mathbf{ \Theta } | \mathcal{H} , \mathcal{I} )$ is the prior on the full set of parameters ($\kappa,\lambda,\mathbf{ \Theta }$), $P( \mathcal{D}|\kappa , \lambda , \mathbf{ \Theta } , \mathcal{H} , \mathcal{I} ) \equiv \mathcal{L}$ is the likelihood, while $P( \mathcal{D} | \mathcal{H}, \mathcal{I} )$ is the model evidence.
The choice of likelihood and prior distributions as well as the numerical implementation of the MCMC technique are discussed further down.

For a given NS--WD system, the gravitational dipole radiation can be described by 3 free parameters, namely the theory parameters ($\kappa, \lambda$) and the central pressure $P_c$ or central density $\rho_c$ of the pulsar, provided that all other orbital details such as the eccentricity, orbital period, etc. are well measured.
The central value of the scalar field $\varphi_c$ on the other hand is obtained through a shooting method from the requirement that the scalar field is zero at infinity $\varphi_{\infty}=0$.

The MCMC runs evolve the 3 parameters through an affine-invariant ensemble sampler available with the Python \verb+emcee+ package \cite{2010CAMCS...5...65G, Foreman-Mackey:2012any}.
The prior distribution is homogeneously distributed in areas of the ($\kappa, \lambda$) space where scalarized solutions exist.
$P_c$ is also sampled homogeneously, but in the range of pressure values which would produce the observed pulsar mass of the system, considering its error bounds.
The dimensionally reduced field equations \eqref{red_field_eq_1} -- \eqref{red_field_eq_4} are solved at each step.
The likelihood function is then computed by comparing the observed decay of the binary system's orbital period with that predicted for the sGB theory with the given parameters from the simulations.
Following \cite{Shao:2017gwu}, the logarithmic likelihood is defined as
\begin{eqnarray}\label{log_likelihood}
    \ln{\mathcal{L}} \propto - \frac{1}{2}\left[ \left( \frac{ \dot{ P }^{\mathrm{int}}_b - \dot{ P }^{\mathrm{th}}_b }{ \sigma^{\mathrm{obs}}_{\dot{P}_b} } \right)^2 + \left( \frac{ m_p/m_c - q }{ \sigma^{\mathrm{obs}}_q } \right)^2 \right],
\end{eqnarray}
where $m_p$ and $m_c$ are the pulsar and companion masses of the system, $q$ is the observed mass ratio, $\dot{P}^{\mathrm{int}}_b$ is the observed intrinsic orbital period decay rate, $\dot{P}^{\mathrm{th}}_b$ is the predicted one and $\sigma^{\mathrm{obs}}_X$ is the uncertainty from observations for the respective quantity $X$.
The second term in \eqref{log_likelihood} is adapted based on whether the ratio $q$ and the WD mass $m_c$ or both masses are known as per Table \ref{table_of_parameters}. 
The predicted orbital decay is dominated by the dipole scalar and quadrupole tensor radiation terms which are given by the following formulae \cite{Damour:1992we, Peters:1964zz} 
\begin{eqnarray}\label{Pdot_dipole}
    \dot{P}^{\mathrm{dipole}}_{b} &=& - \frac{2\pi G_{*}}{c^3} g(e) \left( \frac{2\pi}{P_b} \right) \frac{ m_p m_c }{ m_p + m_c } \left( \frac{D}{m_p} \right)^2, \\ \label{Pdot_quad}
    \dot{P}^{\mathrm{quad}}_{b} &=& - \frac{192\pi G_{*}^{5/3}}{5c^5} f(e) \left( \frac{ 2\pi }{ P_b } \right)^{ 5/3 } \frac{ m_p m_c }{( m_p + m_c )^{1/3} },
\end{eqnarray}
where $m_p$ and $m_c$ are once again the masses of the pulsar and companion respectively, $P_b$ is the orbital period of the binary system, $D$ is the scalar charge of the neutron star in the same units as $m_p$ and $e$ is the eccentricity of the system.
Therefore, for the purpose of the simulation, the period decay is evaluated as $\dot{P}^{\mathrm{th}}_b = \dot{P}^{\mathrm{dipole}}_{b} + \dot{P}^{\mathrm{quad}}_{b}$. 
Lastly, the auxiliary functions of eccentricity $f(e)$ and $g(e)$ are given by
\begin{eqnarray}\label{g_e_def}
    g(e) &\equiv& \left( 1 + \frac{e^2}{2} \right)( 1 - e^2 )^{ -5/2 }, \\ \label{f_e_def}
    f(e) &\equiv& \left( 1 + \frac{ 73 }{ 24 }e^2 + \frac{ 37 }{ 96 }e^4 \right) (1 - e^2)^{ -7/2 }.
\end{eqnarray}

Note that the formulae above \eqref{Pdot_dipole} and \eqref{Pdot_quad} are adopted from the case of standard scalar-tensor theories \cite{Damour:1992we, Peters:1964zz}. The tensor quadrupole emission, eq. \eqref{Pdot_quad}, is modified in sGB gravity, though. However, we can safely use eq. \eqref{Pdot_quad}  since the correction to this standard GR formula due to the Gauss-Bonnet coupling is roughly  of the order $\approx f^\prime (\varphi_0) \frac{mG}{c^2 R} \frac{\lambda^2}{R^2}$ where $\varphi_0$ is the cosmological value of the scalar field, $R$ is the separation between the compact objects and $m$ the total mass of the binary system. In our analysis we have worked with $\varphi_0$ and $ f^\prime (\varphi_0)=0$. But even for $\varphi_0 \ne 0$ this correction is too small in our setting and can be safely neglected. For more details see \cite{Shiralilou:2021mfl}.

Note that the above methodology may also be applied to binary neutron stars (NS--NS pairs).
Those have provided some of the most stringent constraints on modified gravitational theories but have generally been applied to representative EoSs \cite{PhysRevD.54.1474,wex2014testing,Ismail:2017}.
It is significantly more difficult to distinct between the effects of EoS uncertainty and the theory parameters when NS--NS systems are concerned (with one or both NS potentially scalarized).
The use of NS--WD pairs allows us to focus on the variation of parameters for only one of the partners in the system, while considering the WD parameters virtually known and thus constrain the theory parameters better \cite{Berti:2015,Kramer:2016}.

These specific pairs listed in Table \ref{table_of_parameters} have been chosen based on the timing measurement accuracy and the masses (for an excellent summary of the measured pulsar masses until now and tests of general relativity we refer the reader to \cite{FreireWebSite}). As will be shown further down -- the more massive the NS in the pair is -- the better constraints it can provide.  It is worth mentioning that a more massive NS--WD pair than 0348+0432 has already been observed -- J0740+6620 \cite{Fonseca:2021}. Unfortunately there are no direct measurements of its $\dot{P}_b$ value in the literature which can be used to impose constraints as described above. When such observations are made and published, the present work can be extended to obtain further constraint in the theory parameters. The magnitude of the potential improvement is estimated in Sec. \ref{results_1}.

\section{Code setup and numerical results}\label{results}

All numerical simulations are performed with the coupling functions \eqref{eq:coupling} having $\beta = \kappa^{-1}$ as a dimensionless free parameter and for both signs of $\epsilon$.
All results are presented in terms of the dimensionless theory parameter defined as
\begin{eqnarray}
    \lambda \rightarrow \frac{ \lambda }{ R_0 },
\end{eqnarray}
where $R_0 \cong 1.4766$ km corresponds to half the Schwarzschild radius of a 1 solar mass object.
For each integration of the reduced field equations, the boundary conditions \eqref{eq:BC_r0} and \eqref{eq:BC_rinf} are imposed.
Each star is solved as a shooting problem for the central values of $\varphi_c$ and $\Phi_c$ determined by the asymptotics \eqref{eq:BC_rinf}.
The local mass can then be defined through $\Lambda$ as $m = \frac{ r }{ 2 }( 1 - e^{ - 2 \Lambda } )$ and at infinity $m$ gives the total gravitational mass of the neutron star. The scalar charge of the neutron star is obtained from the leading term of the scalar field asymptotics \eqref{eq:Scalar_Charge}
The numerical integration of \eqref{red_field_eq_1}--\eqref{red_field_eq_4} is performed through an adaptive step method based on the Dormand–Prince version \cite{DORMAND198019} of the Runge-Kutta family implemented by the authors.
As was already explored in \cite{Doneva:2017duq}, the shooting procedure is not always easy to converge and some manual work was required to choose appropriate initial guesses for the explored regions of $(\kappa, \lambda)$.
The complete flow of an MCMC run can be understood in terms of the following steps:
\begin{itemize}
    \item Given the ($\kappa, \lambda$, $P_c$) values, a shooting method is performed to obtain the right central value of the scalar field $\varphi_c$ for a NS solution.
    \item The GB neutron star solution is examined for consistency by checking that the asymptotic corresponds to the relationship \eqref{eq:Scalar_Charge}. If so, the scalar charge is extracted.
    \item The dipole and quadrupole radiation terms are computed based on the obtained scalar charge through \eqref{Pdot_dipole}-\eqref{Pdot_quad}, the remaining quantities of the respective system are kept fixed based on the values for the respective NS--WD pair from Table \ref{table_of_parameters}.
    \item The logarithmic likelihood given by eq. \eqref{log_likelihood} is computed using the system measurements, the dipole and quadrupole moments computed in the previous step and the pulsar mass $m_p$ obtained for the given ($\kappa, \lambda$, $P_c$) values. 
    Note that for two of the pulsars $m_p^{\mathrm{obs}}$ is not directly known but derived from the ratio of the NS--WD masses $q \equiv m_p/m_c $.
    \item The computed logarithmic likelihood and the uniform prior functions are passed to the {\tt emcee} package function for evaluation of the posterior distribution and generation of new points.
\end{itemize}

For each of the runs, a total of 416000 points have been used for the {\tt emcee} sampler in a parallel run.
These do not include an initial "shakedown" run of 41600 points used to forget the initial conditions which have been originally chosen uniformly around the prior distribution range.
The {\tt emcee} sampler was run with 32 samplers, meaning that these points are split at 1300 shakedown and 13000 run points per random walker making use of parallel computation with the maximum available machine threads.
Since the $\epsilon = 1$ case turns out to be more difficult numerically (as will be discussed in section \ref{results_2}), we have used twice as many points for obtaining the results there.
While this number did not significantly influence the derived constraints on the theory parameters) (less than 1\% variation), it improved the quality of the plots significantly.

Since there are no obvious limits to the theory parameters ($\kappa, \lambda$), different priors were attempted in order to check the sensitivity of the final constraints. 
The limit $\kappa \rightarrow 0$ translates to $\beta \rightarrow \infty$ that corresponds to GR with a zero scalar field. 
As it turns out, approaching this point leads to undesired numerical problems since we have to be able to calculate scalarized neutron star solutions with almost vanishing scalar field close to the numerical error. 
Furthermore, obtaining scalarized solutions in the large $\kappa$ range is also computationally difficult since the scalarized branches become increasingly short in this case \cite{Doneva:2017duq} and thus solutions exist only  very close to the bifurcation point. 
As shown by the results presented further in the next section \ref{results_0}, this is purely a numerical problem and GB stars are present for arbitrarily large values of $\kappa$. 
Due to these limitation we had to introduce a cutoff of $\kappa$  at both large and small value. The range employed when producing the  figures below is chosen in such a way that the final constrains change by at most 2\% when decreasing/increasing the lower/upped $\kappa$ limit twice. 
This error is much smaller compared to the difference due to different pulsars and EOS.

On the other hand, the $\lambda$ range which provides physically meaningful solutions is highly dependent on the neutron star maximum mass and thus the EoS.
For a given fixed neutron star mass $m_p$, decreasing $\lambda$ for a particular EoS will lead to a critical point $\lambda_{\mathrm{bif}}(m_p)$ for which the bifurcation point from the GR branch is exactly at that NS mass $m_p$.
This value is also dependent on the EoS as we will see and is therefore an important parameter to know for each run since for $ \lambda < \lambda_{\mathrm{bif}}$ the scalarized GB branch does not yet exist and thus the scalar dipole radiation contribution is zero. We should note that beyond the bifurcation point, i.e. $ \lambda > \lambda_{\mathrm{bif}}$ not only the scalarized solutions exist but also the GR ones. 
The latter are both linearly unstable and energetically less favourable, so they can not realize in realistic astrophysical scenarios.

For the $\epsilon = 1$ case, the behaviour is similar but values at which the dipole radiation is within the observational constraints are much tighter.
As described in the conclusion, this shows that for the $\epsilon = 1$ case effectively the $\lambda_{\mathrm{bif}}$ value provides the constraint since scalarized solutions almost immediately exceed the observational bounds with their dipole radiation terms for higher $\lambda$.

\subsection{Interpretation and use of the results -- scalarization with $\epsilon = - 1$}\label{results_0}

The results of each MCMC run have been analyzed and visualized in two different ways: as scatter plots of the sampled points in the free parameters space and as corner plots showing cumulative probability intervals.
The following Fig.\ref{MPA1_Joint_vs_Contour} is an example of these two visualizations for the same run of the MPA1 equation of state (EoS) with the J0348+0432 data and $\epsilon = -1$.

\begin{figure}[ht]
  \subfloat[Scatter Plot]{
	\begin{minipage}[c][1\width]{
	   0.4\textwidth}
	   \centering
	   \includegraphics[width=1\textwidth]{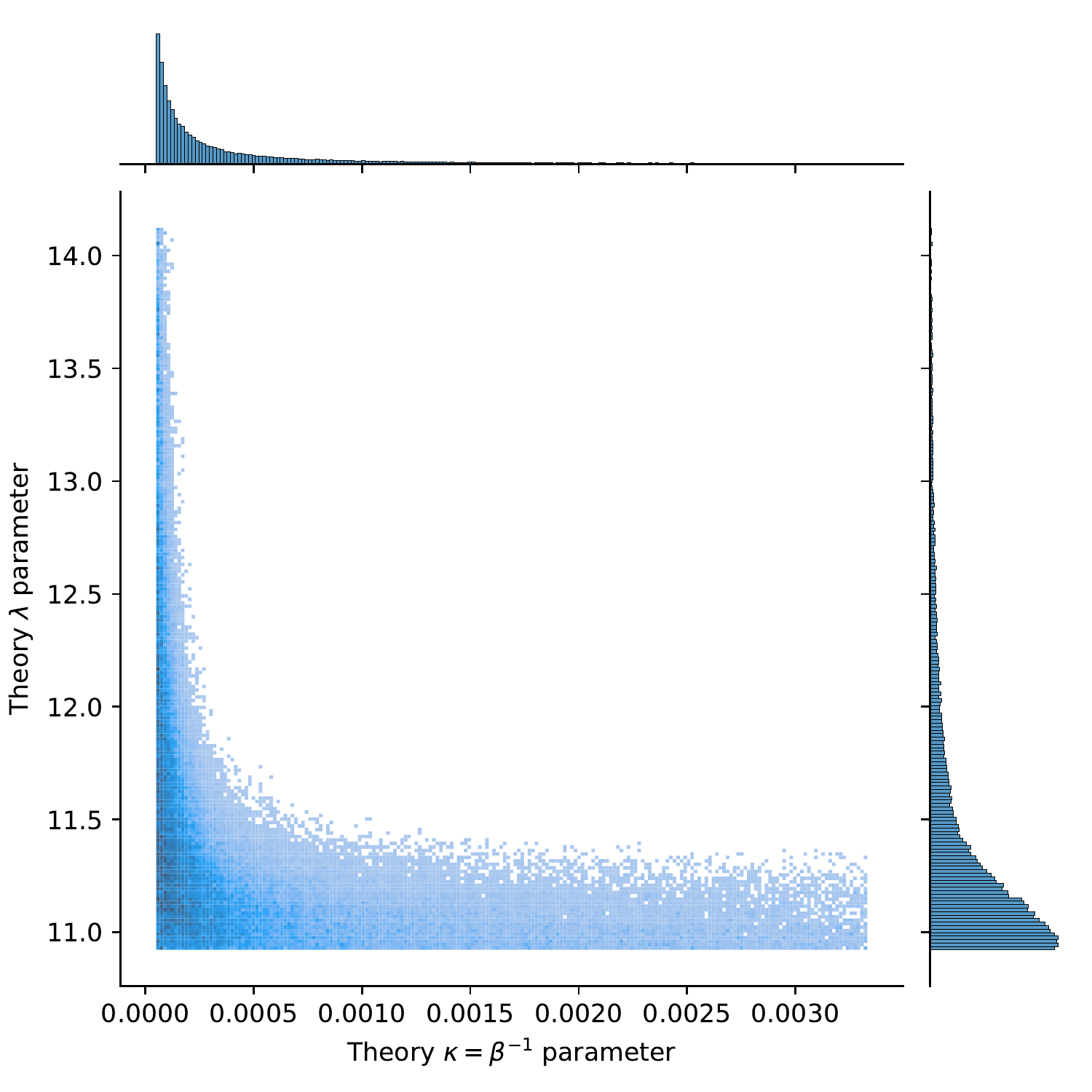}
	\end{minipage}}
  \subfloat[Corner Plot]{
	\begin{minipage}[c][1\width]{
	   0.4\textwidth}
	   \centering
	   \includegraphics[width=1.1\textwidth]{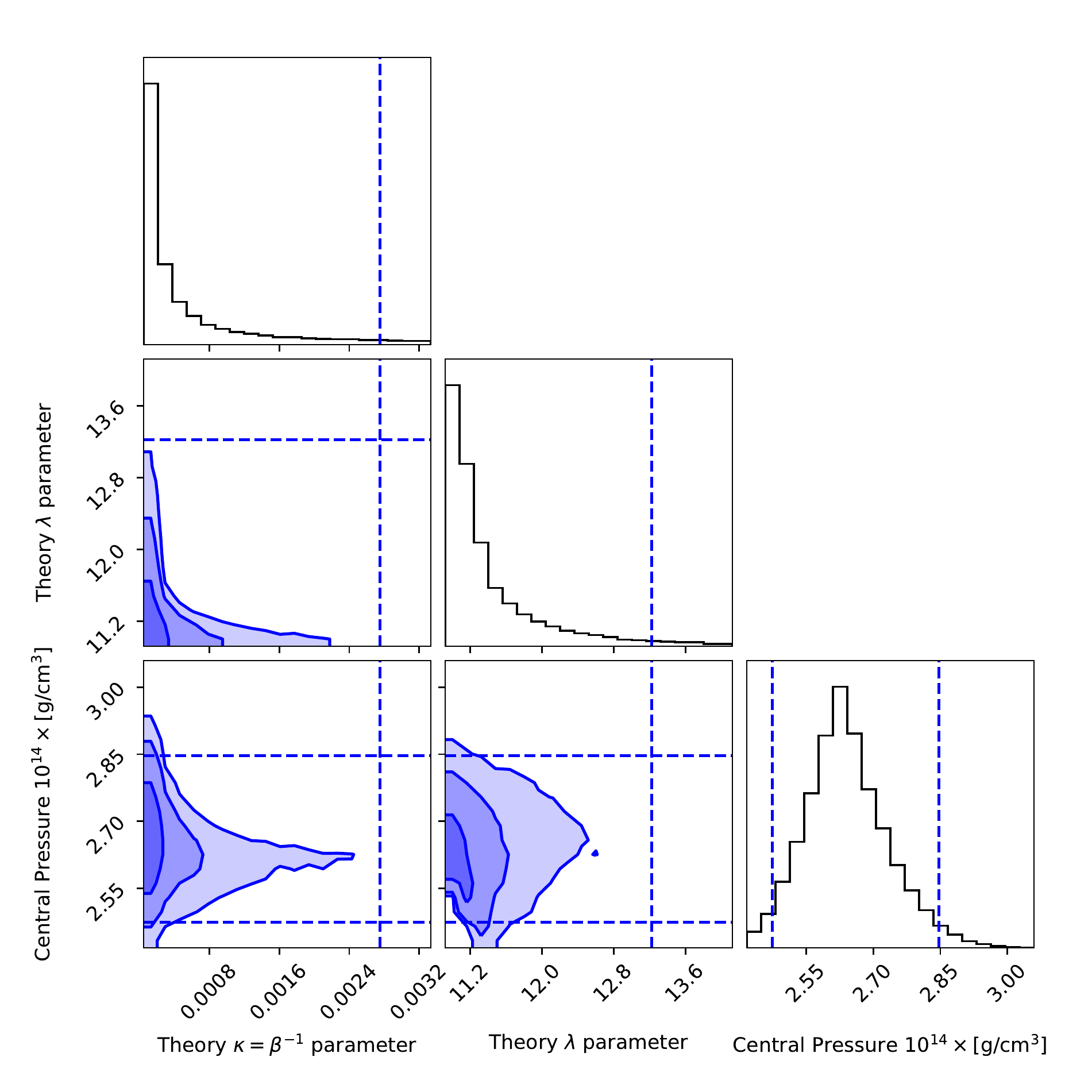}
	\end{minipage}}
\caption{Comparison of the scatter and corner plots for the MPA1 EoS with the J0348+0432 NS--WD pair and $\epsilon = -1$. The scatter plot shows directly the density of samples in the theory parameters space while the corner plot shows the quantiles for 95\%, 80\% and 65\% cumulative probability in the different possible 2D cross-sections. The marginalized 1D distributions for each of the quantities are shown on both subplots as histograms.}\label{MPA1_Joint_vs_Contour}
\end{figure}

The scatter plot shows all points sampled by the MCMC as a 2D cross-section.
This plot is straightforward to understand as the points density distribution in the parameter space.
The higher the density of these points, the higher the posterior probability for that point is.
However, these results cannot be taken directly as a probability.
In addition, the scatter plot shows the marginalized 1-d distributions of the two parameters on the top and right of the main plot.
The Python {\tt seaborn} package \cite{Waskom2021} has been used to generate the scatter plots.

The corner plot on the other hand is a tool used in statistics and in particular MCMC to visualize projections of samples from higher-dimensional distributions.
The lines shown represent quantiles of 95\%, 80\% and 65\% respectively for the posterior, normed within the region of the prior distribution.
The upper right sub-figures in the corner plot show the marginalized 1-d distributions with the 2-sigma levels shown as dashed lines in all 3 subplots.
The Python {\tt corner} package \cite{Foreman-Mackey2016} has been used to generate the corner plots.
The scatter plot gives some insight into the samples from the {\tt emcee} package more directly but the corner plot levels can be interpreted directly as quantiles of the posterior distribution.

The most straightforward way to compare different constraints would be to compare their quantiles with respect to the maximum.
For all EoS and NS--WD pairs analysed, the distributions are very similar in shape and have several common characteristics. 
In all cases the central pressure distribution has a maximum very close to the central pressure for the given pulsar mass and EoS in General Relativity, as expected.
Furthermore, in all cases the $\kappa$ marginalized distributions have similar shapes, decreasing rapidly with higher $\kappa$ and with maximums around $\kappa = 0$.
This fact is easily explained when we note that for lower $\kappa$ the solutions tend closer to GR for an increased interval of $\lambda$ values, making the GB theory indistinguishable from GR in the whole $\lambda$ interval considered for $\kappa < 10^{-4}$.
Indeed, the primary difference between the distributions is the location where the solutions appear ($\lambda_{\mathrm{bif}}$, bellow which we cannot make any inference) and the upper boundary of the $\lambda$ values for the considered range.
Therefore, we have adopted $\lambda_{\mathrm{bif}}$ and the 95\% one-sided quantile interval for $\lambda$ as our constraints for each EoS and each system.

\begin{table}[]
\begin{center}
\begin{tabular}{ |c|c|c|c| } 
 \hline
 NS--WD Pair & $\lambda_{\mathrm{bif}}$ & $\lambda$ 95\% quantile \\
 \hline
 J0348+0432 & 10.92 & 13.22 \\ 
 J1012+5307 & 13.33 & 14.59 \\
 J2222--0137 & 12.47 & 14.12 \\ 
 \hline
\end{tabular}
\caption{Results for the bifurcation line ($\lambda_{\mathrm{bif}}$) and the 95\% one-sided quantile interval for the three pulsars with the MPA1 Equation of State and $\epsilon = -1$.}
\label{res_table_all_PS}
\end{center}
\end{table}

Table \ref{res_table_all_PS} summarizes the results for MPA1 with the three different pulsars for $\epsilon = -1$. 
It is sufficient to demonstrate that the most stringent constraints will arise from the most massive pulsar.
That is why Table \ref{res_table_all_EOS} provides data for the same parameters with different EoS and for J0348+0432 only.
These are plotted and discussed further in the following two subsections.

\begin{table}[]
\begin{center}
\begin{tabular}{ |c|c|c|c| } 
 \hline
 EoS & $\lambda_{\mathrm{bif}}$ & $\lambda$ 95\% quantile \\
 \hline
 MS1 & 16.02 & 17.94 \\ 
 MS1b & 15.79 & 17.73 \\ 
 MPA1 & 10.92 & 13.22 \\ 
 APR3 & 9.63 & 11.98 \\ 
 ENG & 8.72 & 11.11 \\ 
 H4 & 8.60 & 10.83 \\ 
 APR4 & 7.67 & 9.99 \\ 
 WFF2 & 7.38 & 9.94 \\ 
 SLy & 6.78 & 9.40 \\ 
 WFF1 & 6.34 & 8.93 \\ 
 \hline
\end{tabular}
\caption{Results for the bifurcation line ($\lambda_{\mathrm{bif}}$) and the 95\% one-sided quantile interval for the 10 different EoS used with the J0348+0432 NS--WD pair and $\epsilon = -1$.}
\label{res_table_all_EOS}
\end{center}
\end{table}

As will be shown in the follow-up sections, the $\lambda_{\mathrm{bif}}$ is the decisive factor in terms of the constraints an EoS can provide.
For $\lambda < \lambda_{\mathrm{bif}} $ no scalarized GB solutions exist for the given pulsar with mass $m_p$ and thus this region is completely unrestricted by observational data for that pair.

\subsection{Working with different NS-WD pairs -- scalarization with $\epsilon = -1$}\label{results_1}

The following Fig. \ref{MPA1_PS_comparison} shows the scatter density plot for the MPA1 equation of state based on the above discussion for the NS--WD pairs from Table \ref{table_of_parameters}.
For each binary system the dashed line represents the scalarization threshold $\lambda_{\mathrm{bif}}$ while the scatter of points above it reflects the probability density from the MCMC sampler where scalarized neutron stars exist and their scalar radiation is within the observational bounds for the pulsar.
In the region bellow the dashed line scalarization for this pulsar mass is not possible and thus it is by default allowed by the observations of the corresponding binary system.
We have decided not to color the region bellow the dashed line where no constraints can be obtained for the given pulsar in order to avoid too many overlaps.
On the right, a marginalized distribution for each of the pulsars can be seen (integrated over $\kappa$).

As it can be seen, the three double systems do not complement each-other in either case since the ranges of viable constraints have very small overlaps.
While J2222-0137 and J1012+5307 can provide constraints on the parameter space only for $\lambda > 12.5 $, J0348+0432 puts that region at a negligible probability relative to the region $\lambda < 12$.

\begin{figure}[ht]
\centering
\includegraphics[width=0.5\textwidth]{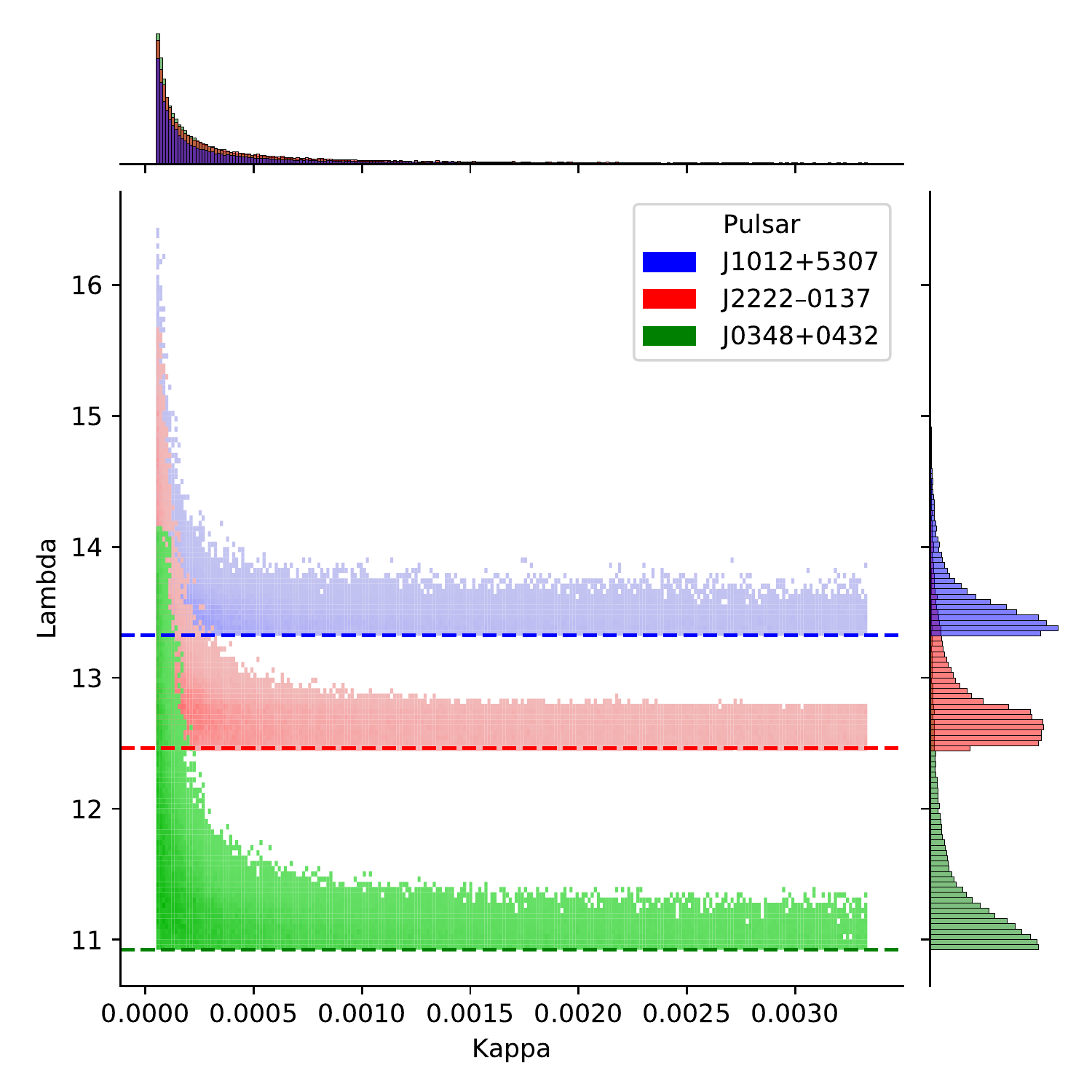}
\caption{Comparison of the bifurcation point $\lambda_{\mathrm{bif}}$ (dashed lines) and the MCMC samples scatter results for the three pulsars listed in Table \ref{table_of_parameters} and the MPA1 equation of state. The colored region for each binary system is where scalarized neutron stars exist and their scalar radiation is within the observational bounds for the given system. The region bellow the dashed line for each system has no constraints for the given NS--WD pair as it is bellow the bifurcation point, i.e. there scalarization for this pulsar mass is not possible and thus the scalar gravitational radiation is zero. Thus, even though this is also a region allowed by observation it has not been colored to avoid too many overlaps.}\label{MPA1_PS_comparison}
\end{figure}

In fact, Fig. \ref{MPA1_PS_comparison} shows that the most stringent constraints from all NS--WD pairs will come from the most massive pulsar J0348+0432 and will not be further constrained even from the inclusion of the remaining two intermediate mass NS-WD binary in Table \ref{table_of_parameters}.
This is the rationale behind presenting results only for a limited number of NS-WD pairs, even though initially more pairs from the list in \cite{FreireWebSite} were considered by us.
We will continue only with the system which has the most massive pulsar J0348+0432 when computing constraints for multiple different EoS.

We should note that an observation of a higher mass pulsar in the future will of course improve these constraints.
The only presently known NS--WD pair which could feasibly provide further constraint is the system J0740+6620 which was already mentioned. 
Based on the NS parameters measured in \cite{Fonseca:2021}, the bifurcation point for it is at $\lambda_{\mathrm{bif}} = 10.26$. 
Once measurements of $\dot{P}_b$ are available, this pair can indeed improve the present constraint by approximately 5\%. 
Taking into account, though, that the present observations support the idea that the neutron star maximum mass is not much higher than two solar masses \cite{Raaijmakers:2019dks,Raaijmakers:2021uju,Nathanail:2021tay}, one can not expect significant improvement beyond this.

\subsection{Constraints from different EoS -- scalarization with $\epsilon = -1$}\label{results_2}

Having demonstrated that the most massive pulsar J0348+0432 gives the strongest constraints, let us turn now to the question how sensitive these constraints are to the nuclear matter EoS.
The following Fig. \ref{PS1_all_eos} outlines the boundaries obtained at the bifurcation point $\lambda = \lambda_{\mathrm{bif}}$ (dashed line) and the MCMC samples scatter results for eight different equations of state.
Only data for the J0348+0432 system is displayed as it provides the most stringent constraints. 
We have used a diverse set of piecewise polytropic approximations of realistic nuclear matter EoS based on \cite{Read:2008iy}. 
These include MPA1, APR3, APR4, ENG, H4, SLy, WFF1, WFF2. In our analysis we included also the MS1 and MS1b EoS that reach higher maximum mass but they provide very loose constraints with $\lambda_{\mathrm{bif}}> 15$ and have been cut out of the figure for better visibility.
The rest of the EoS in \cite{Read:2008iy} have not been used due to their low maximum masses which do not support the existence of a pulsar such as that found in J0348+0432.

\begin{figure}[ht]
\centering
\includegraphics[width=0.5\textwidth]{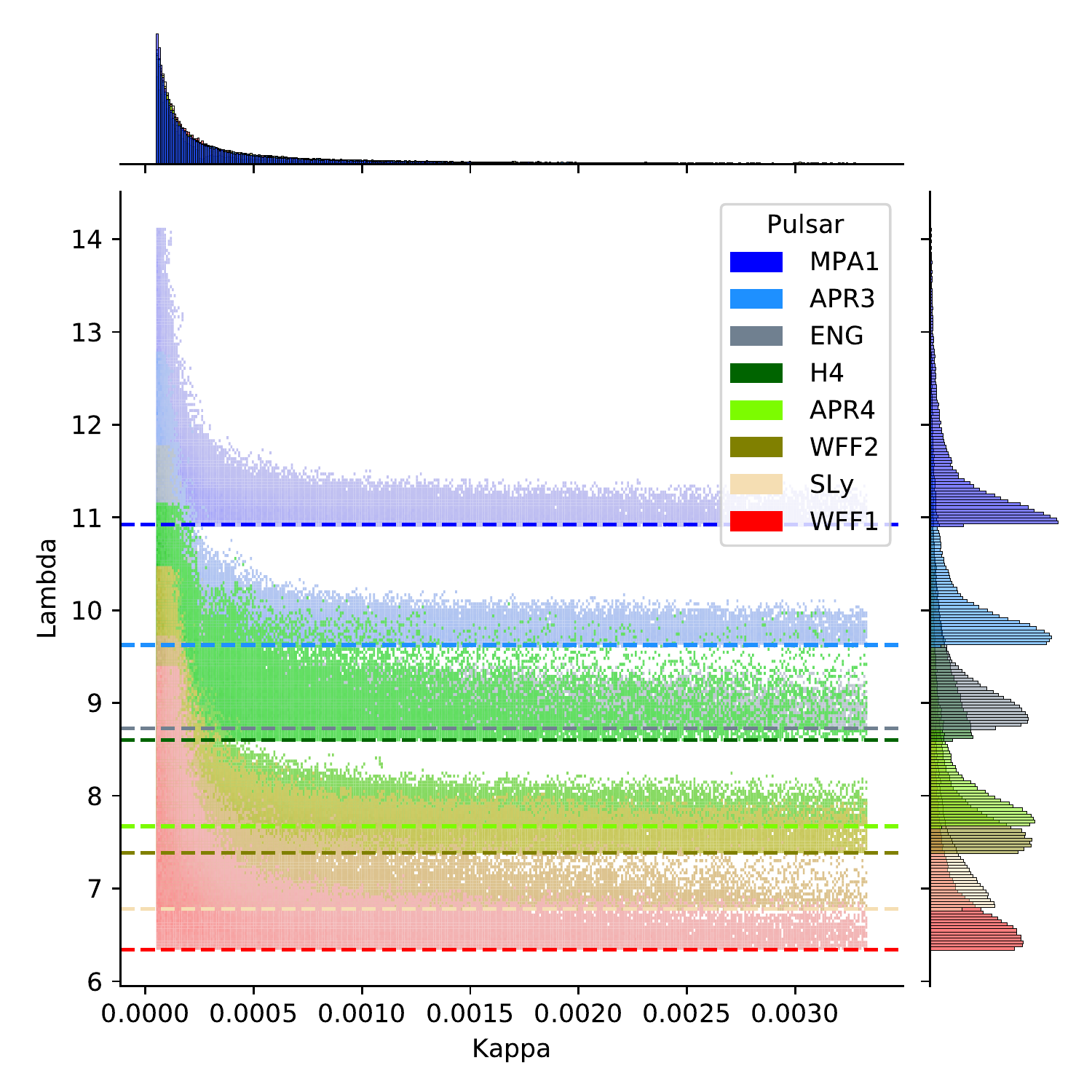}
\caption{ Comparison of the bifurcation point $\lambda_{\mathrm{bif}}$ (dashed line) and the MCMC samples scatter results for J0348+0432 with eight realistic nuclear EoS. The colored region for each EoS is where scalarized neutron stars exist and their scalar radiation is within the observational bounds for the system. The region bellow the dashed line for each EoS has no constraints for the given NS--WD pair as it is bellow the bifurcation point, i.e. there scalarization for this pulsar mass is not possible and thus the scalar gravitational radiation is zero. Thus, even though this is also a region allowed by observation it has not been colored to avoid too many overlaps.}\label{PS1_all_eos}
\end{figure}

It is once again quite evident that except for a few EoS, the domains of mutual constraints are quite limited.
The most stringent limitations on the parameters come from SLy and WFF1 with the rest of the EoS providing virtually no additional constraint.

It is important to note that the regions in $\lambda$-space where probability is non-negligible above the bifurcation point are quite similar for all EOS (as can also be seen in Table \ref{res_table_all_EOS}).
It is therefore evident that the primary factor for the constraints on the $(\lambda,\kappa)$ space is when the bifurcation from GR occurs.
The appearance of this bifurcation, however, has not been unambiguously traced to a EoS trait which can be quantified.
The only observation we have made is that the higher maximum mass EOS, such as MPA1 and APR3 normally lead to weaker constraint.
Taking into account the current EoS constraints \cite{Ozel:2016oaf,Raaijmakers:2019dks,Raaijmakers:2021uju}, as well as the fact that a given EoS should reach the two solar mass barrier, one can safely claim that the constraints shown in Fig. \ref{PS1_all_eos} for the WFF1  EoS can not be significantly improved for another realistic EOS.

\subsection{Constraints with $\epsilon = 1$}\label{results_3}

Unlike the $\epsilon = -1$ case, solutions for the $\epsilon = 1$ coupling function with observationally significant values for the scalar charge are much harder to find in a wide range of the free parameters.
As will be shown on the plots and the results table, the intervals of allowed scalarized solutions above the bifurcation point $\lambda_{\mathrm{bif}}$ are much smaller. 
The reason is that the dipole radiation term for this sign of the coupling function grows to values far exceeding those consistent with the observations at just under 4\% above the bifurcation point $\lambda_{\mathrm{bif}}$.

\begin{figure}[]
  \subfloat[Contour Plot]{
	\begin{minipage}[c][1\width]{
	   0.4\textwidth}
	   \centering
	   \includegraphics[width=1\textwidth]{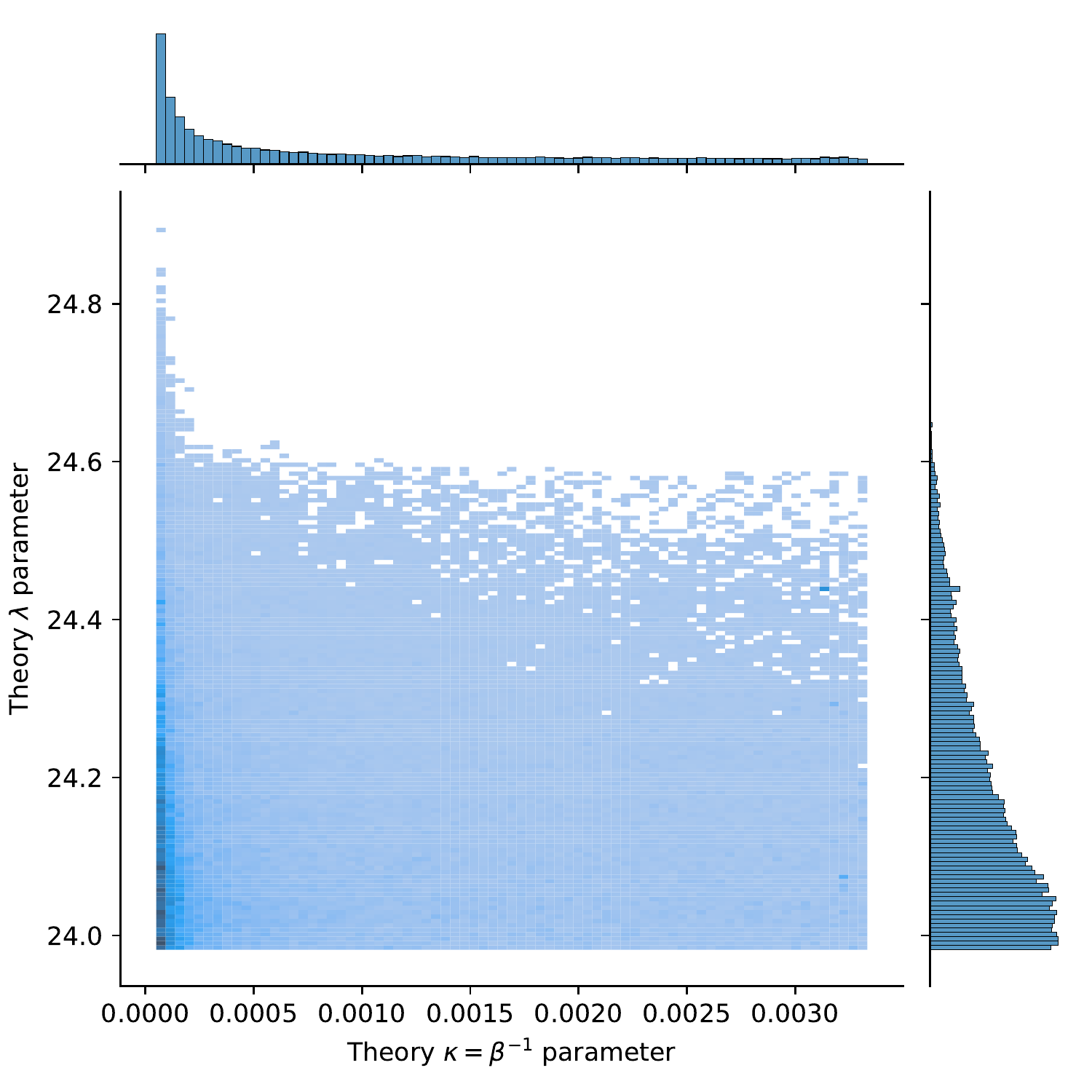}
	\end{minipage}}
  \subfloat[Corner Plot]{
	\begin{minipage}[c][1\width]{
	   0.4\textwidth}
	   \centering
	   \includegraphics[width=1.1\textwidth]{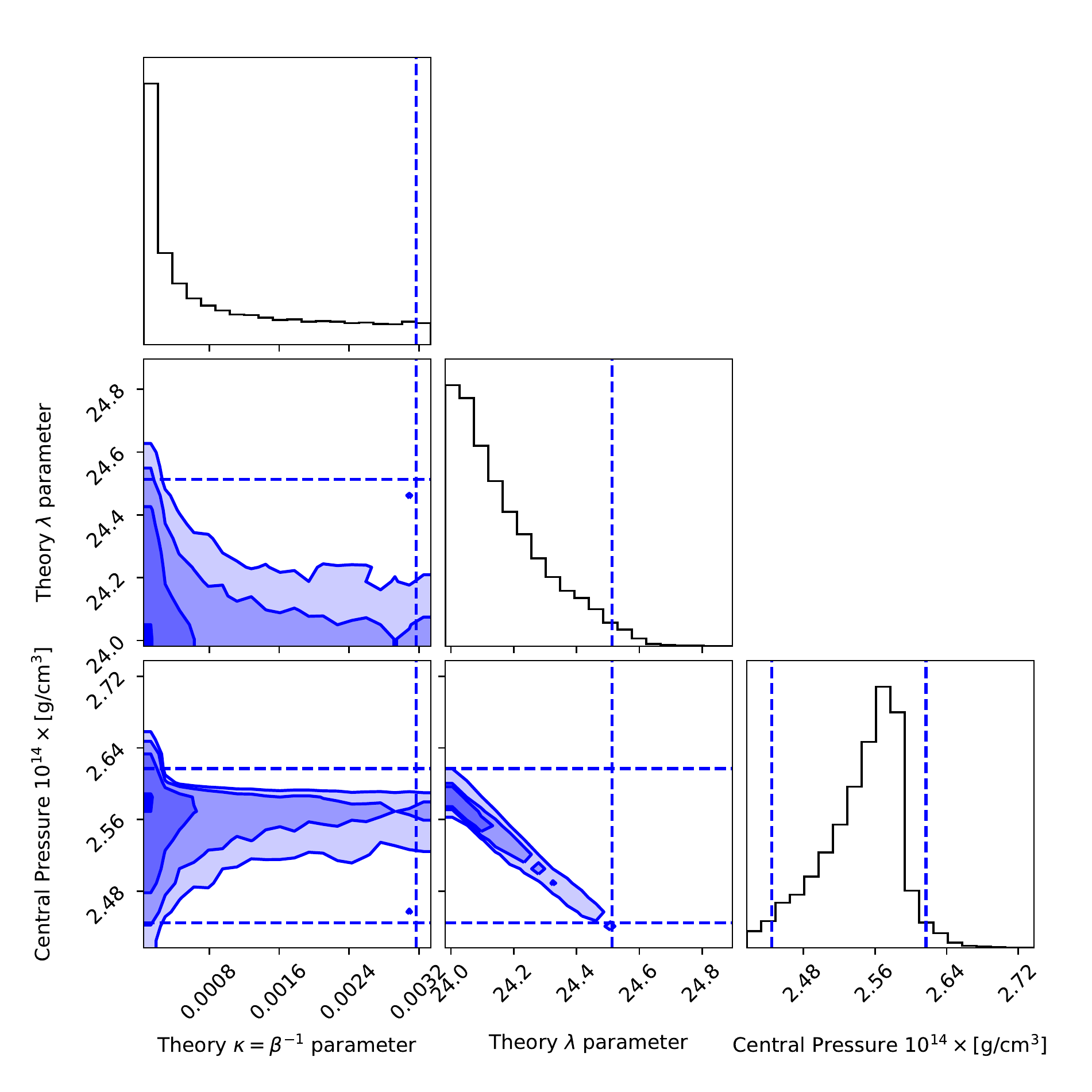}
	\end{minipage}}
\caption{Comparison of the scatter and corner plots for the MPA1 EoS with the J0348+0432 NS--WD pair and $\epsilon = 1$. The scatter plot shows directly the density of samples in the theory parameters space while the corner plot shows the quantiles for 95\%, 80\% and 65\% cumulative probability in the different possible 2D cross-sections. The marginalized 1D distributions for each of the quantities are shown on both subplots as histograms.}\label{MPA1_Joint_vs_Contour_pos_eps}
\end{figure}

The results of the MCMC runs for $\epsilon=1$ have been again analyzed and visualized in two different ways: as scatter and corner plots.
The results are presented in Fig.\ref{MPA1_Joint_vs_Contour_pos_eps} for the MPA1 equation of state (EoS) with the J0348+0432 data.
As one can see on both plots, it is apparent that the scatter of allowed $\lambda$ values is much smaller and the 95\% one-sided quantile interval is very close to the $\lambda_{\mathrm{bif}}$ line (only about 4\% above it).

\begin{figure}[]
\includegraphics[width=0.5\textwidth]{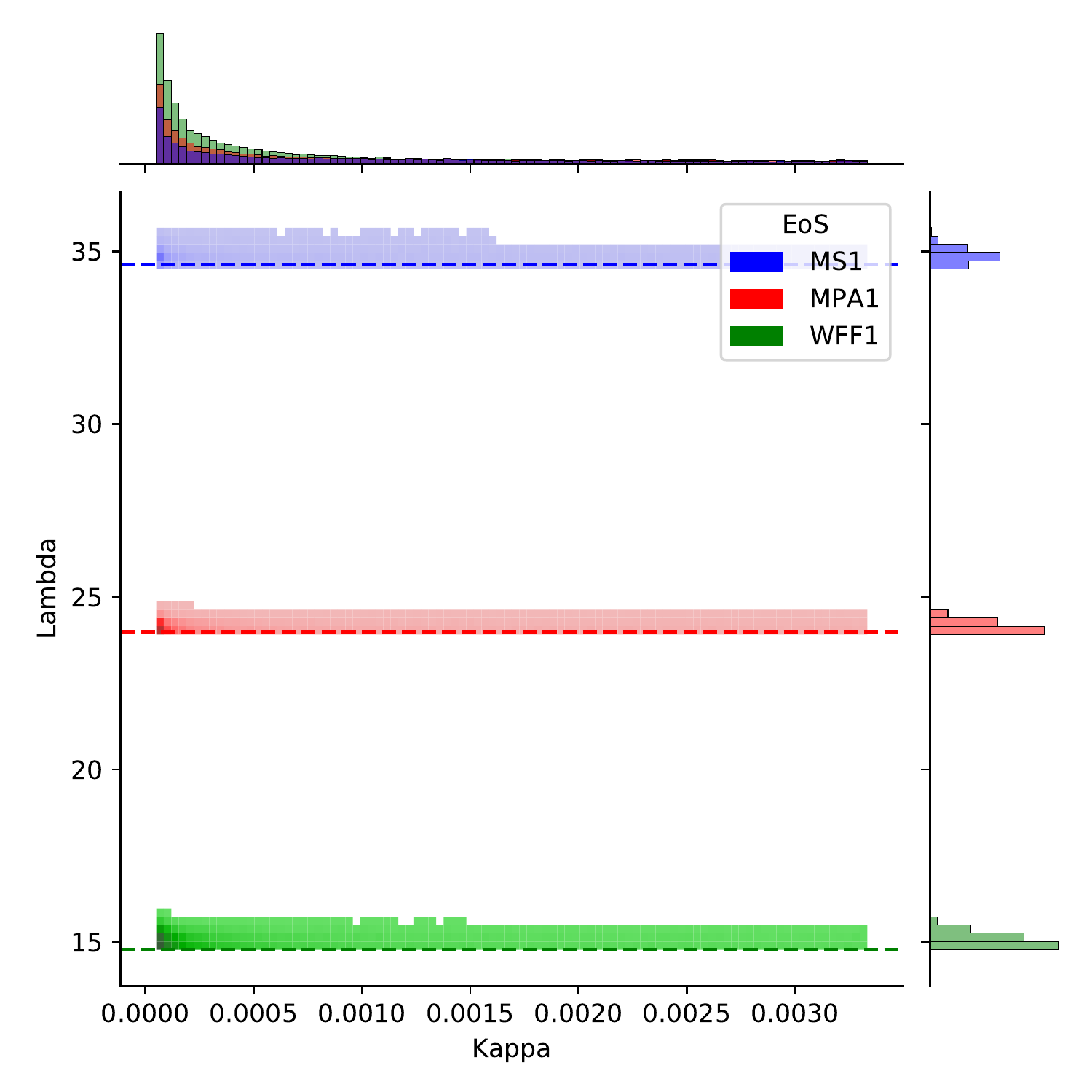}
\caption{Comparison of the bifurcation point $\lambda_{\mathrm{bif}}$ (dashed line) and the MCMC samples scatter results for J0348+0432 with three realistic nuclear EoS.
The colored region for each EoS is where scalarized neutron stars exist and their scalar radiation is within the observational bounds for the system. The region bellow the dashed line for each EoS has no constraints for the given NS--WD pair as it is bellow the bifurcation point, i.e. there scalarization for this pulsar mass is not possible and thus the scalar gravitational radiation is zero. Thus, even though this is also a region allowed by observation it has not been colored to avoid too many overlaps.}\label{PS1_all_eos_pos_eps}
\end{figure}

Fig. \ref{PS1_all_eos_pos_eps} outlines the boundaries obtained at the bifurcation point $\lambda = \lambda_{\mathrm{bif}}$ (dashed line) and the scatter plots resulting from the MCMC runs for the three most relevant equations of state again using only data for the J0348+0432 system that provides the most stringent constraints.
As evident from the figure's marginalized plots on the right, regions of allowed values are much smaller than the $\epsilon = -1$ case in Fig. \ref{PS1_all_eos}. 
The reason for this, as explained above, is that after the bifurcation point the neutron star solutions start developing scalar charge much more rapidly compared to the $\epsilon = -1$ case thus leading to strong scalar dipole radiation that is excluded from the observations.
It should be noted that while it is numerically more difficult for scalarized solutions to converge at larger $\kappa$ (lower $\beta$) with observationally relevant dipole radiation values, they continue to exist for arbitrarily small $\beta$ values.
Therefore, in this case the observational constraint comes from the $\lambda = \lambda_{\mathrm{bif}}$ line directly, since even slightly larger values of $\lambda$ break the observational constraints.

As can be inferred from Table \ref{res_table_all_EOS_poseps}, the values of $\lambda_{\mathrm{bif}}$ and the 95\% one-sided quantile interval for $\lambda$ are within 4\% of each-other.
This shows that the $\epsilon = 1$ case is in fact characterized almost completely by the bifurcation point and there is very little possibility for different EoS complementing each-other in constraining the theory parameters.
On the other hand, the much higher numerical values for $\lambda_{\mathrm{bif}}$ show that this coupling function is less constraining in terms of the GB theory parameters.

\begin{table}[]
\begin{center}
\begin{tabular}{ |c|c|c|c| } 
 \hline
 EoS & $\lambda_{\mathrm{bif}}$ & $\lambda$ 95\% quantile \\
 \hline
 MS1 & 34.51 & 35.35 \\ 
 MPA1 & 23.98 & 24.51 \\ 
 WFF1 & 14.78 & 15.51 \\ 
 \hline
\end{tabular}
\caption{Results for the bifurcation line ($\lambda_{\mathrm{bif}}$) and the 95\% one-sided quantile for the 3 representative EoS used with the J0348+0432 NS--WD pair and $\epsilon = 1$.}
\label{res_table_all_EOS_poseps}
\end{center}
\end{table}

We will continue only with the strongest and weakest constraints coming from WFF1 and MS1 when transferring these constraints to black hole scalarization.

\subsection{Constraining black hole scalarization with $\epsilon=1$}\label{results_4}
Once we have derived and discussed in detail the constraints on the GB theory from binary pulsar observation, it will be interesting to apply them to constrain black hole scalarization.
To do so, we will once again revert to the parametrization in \eqref{eq:coupling} as the more standard one.
For fixed $\lambda$ and $\beta$, though, a whole branch of scalarized black holes can exist spanning from the bifurcation point to zero black hole mass. 
We have decided to focus on two characteristics features of these branches. 
The first one is the black hole mass at the point of bifurcation that represents the maximum mass of a scalarized black holes and depends only on $\lambda$. 
The second one is the maximum scalar charge that can be reached for the whole sequence of scalarized solutions with fixed $\lambda$ and $\beta$ that is directly connected to the emitted scalar dipole radiation during black hole inspiral for example.

The following Fig. \ref{PS1_BH_constraints} shows the constraints from Table \ref{res_table_all_EOS_poseps} overlaid on top of the maximum scalar charge and the maximum black hole mass for scalarized black holes in the sGB theory with corresponding ($\beta, \lambda$) value.
We have displayed the constraints only for two EoSs, namely WFF1 and MS1, that represent the two limiting cases considered in the previous subsection with the strongest and the weakest constraint respectively.  
\begin{figure}[ht]
  \subfloat[Maximum BH mass in $\mathrm{M}_{\odot}$]{
	\begin{minipage}[c][1\width]{
	   0.5\textwidth}
	   \centering
	   \includegraphics[width=1\textwidth]{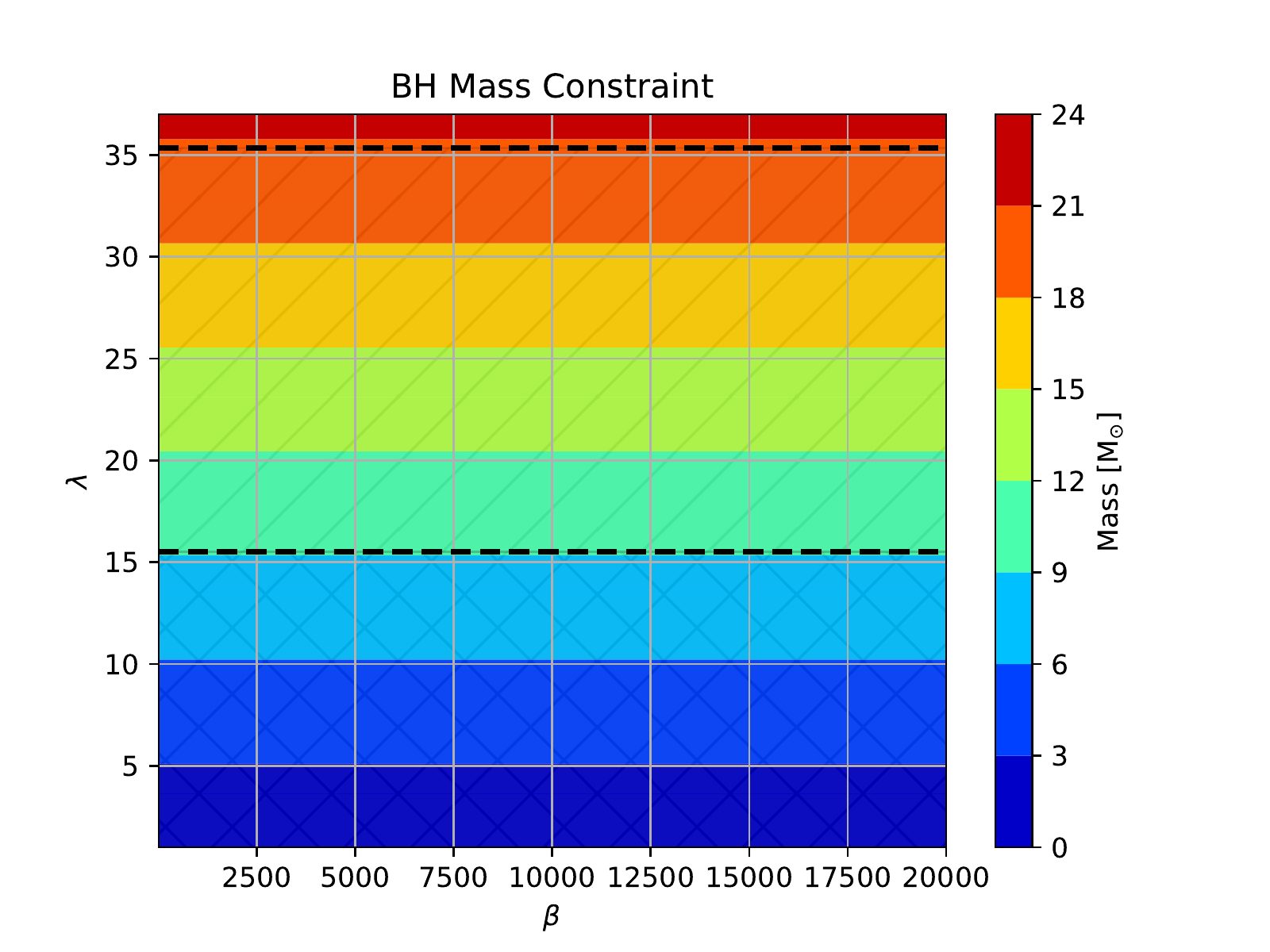}
	\end{minipage}}
   \subfloat[Natural log of the maximum BH scalar charge.]{
	\begin{minipage}[c][1\width]{
	   0.5\textwidth}
	   \centering
	   \includegraphics[width=1\textwidth]{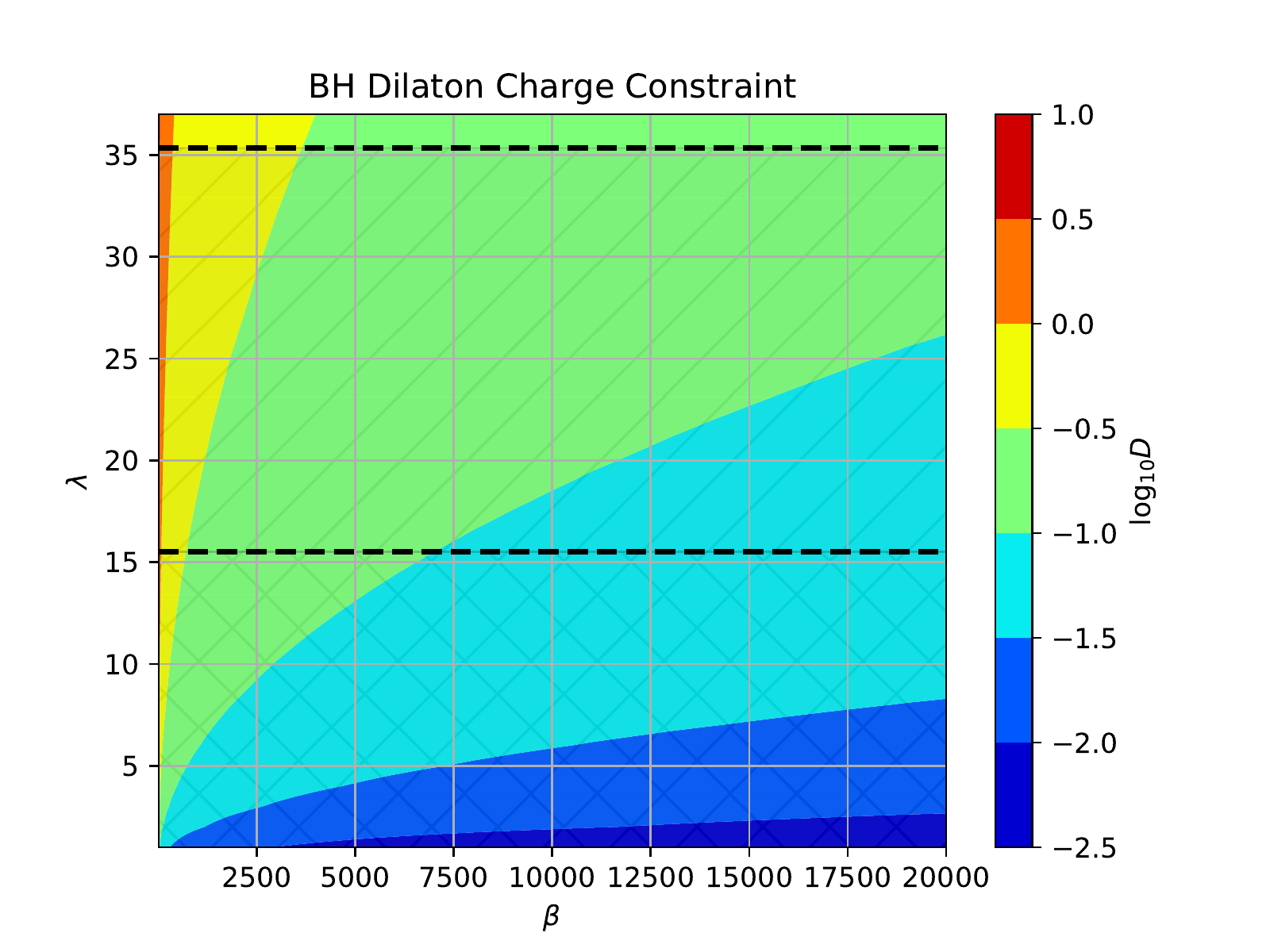}
	\end{minipage}}
 \caption{ Comparison of the lowest and highest constraints arising from the J0348+0432 system based on the different EoS used with 95\% one-sided quantile interval for $\lambda$. The shaded region with ``x'' pattern is where none of the EoSs provide any constraints on the parameters, i.e. scalarized neutron stars do not exist. The region marked with "/" is where at least one EoS provides some constraint on the parameters, i.e. scalarization exist for at least one of the considered EoS. The unmarked region is the one excluded from observations. 
 }\label{PS1_BH_constraints}
\end{figure}

Let us focus first on the maximum mass of scalarized black holes depicted in the left panel of Fig. \ref{PS1_BH_constraints}. This is the mass taken at the point of bifurcation and it does not dependent on $\beta$ but scales with $\lambda$ \cite{Doneva:2017bvd,Doneva:2018rou}. As one can see, the maximum mass that is still in agreement with binary pulsar observations is roughly $20M_\odot$ and this is for the relatively extreme MS1 EoS. If one takes a more conservative choice of an EoS that is in the preferred observational range \cite{Ozel:2016oaf,Raaijmakers:2019dks,Raaijmakers:2021uju}, this constraints can easily drop to roughly $9 M_\odot$. 
This practically excludes the possibility for scalarization for at least two thirds of the binary black hole mergers observed until now through GW.
Thus, even if scalarized black holes in sGB gravity exist they can account only for a limited number of GW events.

The black hole scalar charge on the other hand can reach relatively large values for the observationally allowed range of $\lambda$ and $\beta$, as seen in the  right panel of Fig. \ref{PS1_BH_constraints}. As the results in \cite{Maselli:2020zgv,Maselli:2021men} show, such values of $D$ can be tested also with EMRIs, assuming that the small orbiting object is a scalarized black hole. Still, the problem remains that for a big portion of observed EMRIs, the orbiting black hole might have mass larger than the bounds discussed above, and thus not scalarize.

Here we have discussed the constraints on black hole scalarization only for  $\epsilon=1$. The reason is that for  $\epsilon=-1$ only rotating black holes can develop nontrivial scalar hair and a slightly different phenomenon is present which is the so-called spin-induced scalarization.
The essence is that for rapidly rotating black holes the GB invariant can change sign close to the black hole horizon thus allowing for scalarization with opposite sign of the coupling function \cite{Dima:2020yac,Hod:2020jjy,Doneva:2020nbb,Berti:2020kgk,Herdeiro:2020wei,Doneva:2020kfv}.
Here we will not transfer the binary pulsar constraints with $\epsilon=-1$ to black holes because of the complexity of finding rotating black hole solutions.
In addition, we will have one more parameter -- the black hole angular momentum, that makes imposing constraints a formidable task.
Thus we leave this for future studies.

\section{Conclusions}\label{conclussions}
Pulsars in close binary systems have proven over the years to give one of the most stringent constraints on strong field regime of gravity. The perfect fit of the pulsar's orbit shrinking due to gravitational wave emission against the GR predictions has already put strong limits on a number of theories predicting an additional channel of energy loss such as scalar dipole radiation. In the present paper we have focused on a quite popular class of alternative theories of gravity that has not been studied yet in the context of binary pulsars, namely scalar-Gauss-Bonnet gravity admitting scalarization. The large interest in this theory comes from the fact that it offers an elegant way to produce black holes with scalar hair while leaving the weak field regime of gravity unaltered. Even though different astrophysical implications of such scalarized black holes were considered in the literature, the strong constraints coming from binary pulsar experiment have not been explored yet. Our goal was to fill this gap by considering a variety of observed double systems in the context of sGB gravity.   

More specifically we have considered the three most suitable pairs of NS-WD binaries and examined the constraints these systems can impose on the theory parameters $\lambda$ and $\beta$ through Markov Chain Monte-Carlo (MCMC) methods applied to the observations. Here $\lambda$ is the coupling constant between the Gauss-Bonnet invariant and the scalar field and $\beta$ is a coefficient in the coupling function. We have considered both positive and negative signs of the coupling function. While neutron stars can be scalarized independent of this sign, standard and spin-induced scalarization can be distinguished for black holes that leads to quite distinct properties of the hairy black holes.

For a given pulsar one can determine a minimum $\lambda_{\mathrm{bif}}$ bellow which scalarization is not possible (independent of $\beta$) and thus for such values of the parameters the sGB theory is by definition in agreement with the binary pulsar observations. For $\lambda>\lambda_{\mathrm{bif}}$, though, there is a range in the $\lambda-\beta$ plane where scalarization is allowed while the scalar dipole radiation is small enough so that it is still within the observational uncertainty and is thus allowed by observations. We should point out, though, that this region with $\lambda>\lambda_{\mathrm{bif}}$ is relatively narrow, especially for a positive sign of the coupling function (the sign that leads to scalarization of both static and rotating black holes).

The results show that the more massive a pulsar is, the stronger the constraints on sGB gravity. In addition, these constraints are of course dependent on the nuclear matter EOS we employ but if we limit ourselves to modern realistic equations of state that are in agreement with the astrophysical observations, the spread is relatively small. With the discovery of new pulsars in close binary systems and the refinement of our knowledge of the high density equation of state, the constraints we derive in the present paper might of course be improved. Based on our results, though, and the present astrophysical observations related to the maximum neutron stars mass one can conclude that the expected improvement can be up to a few tens of percent at most. One should keep in mind that these results are for sGB gravity with a massless scalar field. For a nonzero mass, the scalar charge will be zero and the scalar gravitational radiation will be suppressed similar to the DEF model \cite{Ramazanoglu:2016kul,Yazadjiev:2016pcb} thus leading to agreement with binary pulsar observations for more or less arbitrary $\lambda$ and $\beta$. 

Having derived constraints on the sGB parameters $\lambda$ and $\beta$, we have transferred them to static scalarized black holes in order to explore the possible range of astrophysically relevant solutions with scalar hair. We have focused on two characteristics of the scalarized black holes -- the maximum mass a scalarized black hole can achieve and the maximum scalar charge. We have obtained that the maximum mass of a static scalarized black hole is roughly $20M_\odot$ if we consider a broad set of equations of state, but drops to approximately $9M_\odot$ for equations of state that are in the preferred range according to astrophysical observations such as NICER and the binary neutron star mergers. Thus, it turns out that it is not that easy to observe a scalarized black hole because of the fact that it should have a relatively small mass and only a  limited number of such objects are expected to be observed in the near future. The maximum black hole scalar charge that is allowed by the binary pulsar observations is relatively large, though. Thus it can lead to clear observational signatures. These results were obtained for one form of the coupling function that is quite generic. Based on the experience in other related problems, we believe that the constraints on the black hole properties will not change considerably even if we modify the coupling while still allowing for scalarization. This is a study underway.
 
\begin{acknowledgments}
We would like to thank the anonymous referee for helpful suggestions in improving the manuscript. We would also like to thank P.~Freire for helpful commends and suggestions. VD, DD, and SY acknowledge the support by the National Science Fund of Bulgaria under Contract No. K-06-N-38/12. DD acknowledge financial support via an Emmy Noether Research Group funded by the German Research Foundation (DFG) under grant no. DO 1771/1-1. SY would like to thank the University of T\"ubingen for the financial support. Networking support by the COST actions CA15117 and CA16104 is gratefully acknowledged.
\end{acknowledgments}

\bibliography{references}

\end{document}